\documentclass[aps,reprint,superscriptaddress,nofootinbib]{revtex4-2}
\usepackage{sidecap}

\usepackage{ulem}
\usepackage{epsfig}
\usepackage{amsmath,amssymb,amsthm,mathtools}
\usepackage{graphicx}
\usepackage{hyperref}
\usepackage{bm,url,enumitem}
\usepackage{comment}
\usepackage{color,soul}
\setstcolor{blue}
\sethlcolor{blue}
\usepackage{xcolor}
\DeclareMathAlphabet{\mathpzc}{OT1}{pzc}{m}{it}

\usepackage{sidecap}
\usepackage{amsmath}
\usepackage{ulem}
\usepackage{epsfig}
\usepackage{amsmath,amssymb,amsthm,mathtools}
\usepackage{graphicx}
\usepackage{hyperref}
\usepackage{bm,url,enumitem}
\usepackage{comment}
\usepackage{color,soul}
\setstcolor{blue}
\sethlcolor{blue}
\usepackage{xcolor}

\begin{document}

\newcommand{\vAi}{{\cal A}_{i_1\cdots i_n}} \newcommand{\vAim}{{\cal
A}_{i_1\cdots i_{n-1}}} \newcommand{\vAbi}{\bar{\cal A}^{i_1\cdots i_n}}
\newcommand{\vAbim}{\bar{\cal A}^{i_1\cdots i_{n-1}}}
\newcommand{\htS}{\hat{S}} \newcommand{\htR}{\hat{R}}
\newcommand{\htB}{\hat{B}} \newcommand{\htD}{\hat{D}}
\newcommand{\htV}{\hat{V}} \newcommand{\cT}{{\cal T}} \newcommand{\cM}{{\cal
M}} \newcommand{\cMs}{{\cal M}^*}
 \newcommand{\vk}{{\bf k}}
\newcommand{\vK}{{\vec K}} \newcommand{\vb}{{\vec b}} \newcommand{{\vp}}{{\vec
p}} \newcommand{{\vq}}{{\vec q}} \newcommand{\vQ}{{\vec Q}}
\newcommand{\vx}{{\vec x}}
\newcommand{\tr}{{{\rm Tr}}} 
\newcommand{\beq}{\begin{equation}}
\newcommand{\eeq}[1]{\label{#1} \end{equation}} 
\newcommand{\half}{{\textstyle
\frac{1}{2}}} \newcommand{\gton}{\stackrel{>}{\sim}}
\newcommand{\lton}{\mathrel{\lower.9ex \hbox{$\stackrel{\displaystyle
<}{\sim}$}}} \newcommand{\ee}{\end{equation}}
\newcommand{\ben}{\begin{enumerate}} \newcommand{\een}{\end{enumerate}}
\newcommand{\bit}{\begin{itemize}} \newcommand{\eit}{\end{itemize}}
\newcommand{\bc}{\begin{center}} \newcommand{\ec}{\end{center}}
\newcommand{\bea}{\begin{eqnarray}} \newcommand{\eea}{\end{eqnarray}}
\newcommand{\beqar}{\begin{eqnarray}} \newcommand{\eeqar}[1]{\label{#1}
\end{eqnarray}} \newcommand{\bra}[1]{\langle {#1}|}
\newcommand{\ket}[1]{|{#1}\rangle}
\newcommand{\norm}[2]{\langle{#1}|{#2}\rangle}
\newcommand{\brac}[3]{\langle{#1}|{#2}|{#3}\rangle} \newcommand{\hilb}{{\cal
H}} \newcommand{\pleft}{\stackrel{\leftarrow}{\partial}}
\newcommand{\pright}{\stackrel{\rightarrow}{\partial}}

\title{Cone-Dependent Jet Collisional Energy Loss in Finite QCD Medium}

\author{Magdalena Djordjevic}
\email{magda@ipb.ac.rs}
\affiliation{Institute of Physics Belgrade, University of Belgrade, 11080 Belgrade, Serbia}
\affiliation{Serbian Academy of Sciences and Arts, 11000 Belgrade, Serbia}

\author{Bojana Ilic}
\affiliation{Institute of Physics Belgrade, University of Belgrade, 11080 Belgrade, Serbia}

\author{Marko Djordjevic}
\affiliation{Faculty of Biology, University of Belgrade, 11000 Belgrade, Serbia}

\begin{abstract}
We derive a compact HTL-resummed expression for the leading-order jet collisional energy loss in a finite-size, finite-temperature QCD medium. Defining the jet energy inside a cone of radius $R$, we obtain the out-of-cone elastic energy loss with an explicit separation between contributions from the primary jet parton and recoiling medium partons. The result reproduces the known partonic limit as $R\!\to\!0$, vanishes for $R\!\to\!\pi$, and applies to both light- and heavy-flavor jets.
Numerically, the elastic component shows a pronounced non-linear $R$ dependence relative to the radiative baseline, and its importance increases with $R$, becoming comparable to or exceeding the radiative contribution for sufficiently large jet radii. The path-length dependence remains close to linear for all $R$, while the medium-response contribution can exceed $10\%$ for realistic jet radii.
\end{abstract}

\maketitle

\section{Introduction}

The suppression of high-$p_T$ hadrons and  jets in heavy-ion collisions is commonly attributed to the energy loss of energetic partons as they traverse the deconfined quark--gluon plasma (QGP). Medium-induced gluon radiation has long been regarded as the dominant mechanism, and the seminal perturbative formalisms (e.g. BDMPS/ASW~\cite{BDMPS,Z,ASW}, (D)GLV~\cite{GLV,DG}, higher-twist~\cite{HT,HTM}, AMY~\cite{AMY}) were all originally formulated in terms of radiative energy loss only. Elastic (collisional) processes were either neglected or estimated to be subleading. 

Subsequent comparisons with data and systematic heavy-flavor studies have made clear that collisional energy loss is not negligible. In~\cite{Wicks2007,Mustafa} it was shown that elastic energy loss is of the same order as radiative loss for realistic QGP parameters and may even dominate for bottom quarks. Studies based on modified Langevin dynamics indicate that collisional processes can be significant at low and intermediate transverse momenta and remain comparable to gluon radiation over a broad momentum range~\cite{Cao2013}.
 Including elastic interactions improves the description of heavy-flavor
suppression and modifies the $p_T$-dependence of the nuclear modification factor $R_{AA}$, underscoring the need to specify how collisions are treated and how the transferred energy is redistributed~\cite{Blagojevic:2014a,Cao:2017zih}.

In Ref.~\cite{Djordjevic2006}, a finite-size HTL-resummed expression for the leading-order
(partonic) collisional energy loss was derived in a dynamical QCD medium.
While that formalism is well suited for parton-level energy loss, current jet-quenching phenomenology
is increasingly driven by reconstructed-jet observables and their dependence on the jet radius $R$~\cite{Cao:2024JetQuenching,Barreto:2025JetConeRadius,ALICE:JetR2024}, which directly probes angular redistribution and in-cone energy recovery. In particular, the $R$-dependence of inclusive jet suppression is now directly constrained by LHC measurements over a broad range of jet radii~\cite{ALICE:JetR2024,ATLAS:DijetR2024}  and has been emphasized as a sensitive handle on the role of recoil and medium response (see, e.g., Refs.~\cite{Wang2017_recoil,MehtarTani:2021dqn}). This motivates extending the finite-size collisional framework to jets by explicitly accounting for the energy carried by recoiling medium partons relative to a finite cone, beyond limiting assumptions where the transferred energy is either fully lost outside the cone or rapidly thermalized into the medium.

Existing jet-quenching models implement elastic energy transfer and recoil with different assumptions. In several widely used frameworks, collisional energy loss is approximated as a drag/diffusion acting on the leading parton and soft recoils are absorbed into the medium background (e.g.\ MARTINI~\cite{Schenke2009}); PYQUEN/HYDJET++ treats the collisional component as an explicit out-of-cone loss~\cite{Lokhtin2011}; and hybrid strong/weak-coupling approaches assume rapid thermalization of the deposited energy without tracking recoils~\cite{Casalderrey2014_hybrid,Casalderrey2017_hybridAngular}.
More recent transport and Monte-Carlo frameworks propagate recoil partons and model medium response explicitly (e.g.\ LBT and JEWEL with recoils)~\cite{He2015_LBT,Luo2024_LBTinel,Elayavalli2017_JEWELrecoil,Milhano2018_substructure}, but necessarily rely on model-dependent implementations rather than a unified microscopic derivation.

In this work we extend the finite-size HTL formalism of Ref.~\cite{Djordjevic2006} from a single energetic parton to a jet by defining the jet energy inside a cone of radius $R$ and deriving the corresponding cone-restricted collisional energy loss from the same one-loop diagram. The resulting analytic expression separates the primary-parton and recoil contributions, interpolates between the partonic limit ($R\!\to\!0$) and complete recovery ($R\!\to\!\pi$), and applies to both light- and heavy-flavor jets.
We then study its dependence on $R$, jet energy and mass, path length, and the medium-response contribution associated with the $\omega<0$ sector.

\section{Formalism}

In this section, we follow the notation and conventions of Ref.~\cite{Djordjevic2006} and present only the steps needed for the cone-restricted generalization (for more details, see Supplementary Material~\cite{CollisionalJetSupp}). We first recall the main ingredients needed for the present derivation and refer the reader to Ref.~\cite{Djordjevic2006} for further details. We work in
Coulomb gauge, where the HTL--resummed gluon propagator can be written as
\begin{equation}
  D^{\mu\nu}(\omega,\mathbf{q})
  = - P^{\mu\nu}\,\Delta_T(\omega,\vec{\mathbf{q}})
    - Q^{\mu\nu}\,\Delta_L(\omega,\vec{\mathbf{q}}),
  \label{eq:HTL_propagator}
\end{equation}
with the transverse and longitudinal propagators
\beq
\Delta_T^{-1} = \omega^2 - \vec{\mathbf{q}}^{2} - \frac{\mu_E^{2}}{2} -
\frac{(\omega ^{2} - \vec{\mathbf{q}}^{2})\mu_E^{2}}{2 \vec{\mathbf{q}}^{2}} 
(1+\frac{\omega }{2|\vec{\mathbf{q}}|}
\ln |\frac{\omega -|\vec{\mathbf{q}}|}{\omega +|\vec{\mathbf{q}}|}|),
\eeq{DeltaT}
\beq
\Delta_{L}^{-1}= \vec{\mathbf{q}}^{2}+ \mu_E^{2}
(1+\frac{\omega }{2|\vec{\mathbf{q}}|} 
\ln |\frac{\omega -|\vec{\mathbf{q}}|}{\omega +|\vec{\mathbf{q}}|}|),
\eeq{DeltaL}
and the only nonzero
components of the projectors
\begin{equation}
  P_{ij} = \delta_{ij} - \frac{q_i q_j}{\vec{\mathbf{q}}^2},
  \qquad Q_{00} = 1.
\end{equation}

\subsection{Finite--size jet collisional energy loss}
We consider the $0^{th}$ order in opacity diagram with one HTL gluon exchanged
between the energetic parton and a thermal medium parton. Requiring that the
interaction occur inside a medium of size $L$ leads to the finite-size factor in the elastic amplitude,
\begin{equation}
  i M_{\rm el}
  = J(p')\,\frac{1}{2E}\,
    \frac{1 - e^{-i(E-E'-\omega)L/v}}{E-E'-\omega}\; i{\cal M},
  \label{eq:Mel}
\end{equation}
where 
{\small
\beqar
{\cal M}=g^2 
D_{\mu \nu } (k^\prime-k) \bar{u} (p^\prime, s^\prime) \gamma^\mu u (p, s) 
\bar{u} (k^\prime, \lambda^\prime) \gamma^\nu u (k, \lambda) \hspace{1.2em}
\eeqar{M_BT}} 
\hspace{-0.8em} and the full derivation of this finite-size factor is given in
Ref.~\cite{Djordjevic2006}. Here $E=\sqrt{M^2+\vec{\mathbf{p}}^2}$, $M$ is the quark mass, 
$\vec{\mathbf{p}}= \vec{\mathbf{p^\prime}}+(\vec{\mathbf{k^\prime}}-\vec{\mathbf{k}})$ and 
$\omega=|\vec{\mathbf{k^\prime}}|-|\vec{\mathbf{k}}|$ is the energy transfer from the medium parton. For highly energetic jets, 
where $|\vec{\mathbf{q}}| \ll E$, $E^\prime$ becomes 
$E^\prime \approx E-\vec{\mathbf{v}}\cdot\vec{\mathbf{q}}$. Here 
$\vec{\mathbf{v}} =  \mathbf{p}/E$ is the velocity of the initial jet, i.e., the jet
4-momentum $p$ is equal to 
$p=\left(\frac{M}{\sqrt{1-v^2}}, \frac{M \vec{\mathbf{v}}}{\sqrt{1-v^2}}\right)$.
Squaring the amplitude and summing over spins yields
{\small \begin{equation}
  \frac{1}{2}\sum_{\rm spins} |M_{\rm el}|^2
  = |J(p')|^2\,\frac{1}{E^2}\,
    \frac{\sin^2\left((\omega-\vec{\mathbf{v}}\cdot\vec{\mathbf{q}})\frac{L}{2v}\right)}
         {(\omega-\vec{\mathbf{v}}\cdot\vec{\mathbf{q}})^{2}}\,
    \frac{1}{2}\sum_{\rm spins} |{\cal M}|^2,
  \label{eq:Mel2}
\end{equation}}
\hspace{-1.3em} where $\sum_{\rm spins} |{\cal M}|^2$ is given by Eq.~\ref{Cal_M_Sq} below (see also~\cite{Djordjevic2006}).

The differential energy loss is given by $dE_{el}= \omega \,d\Gamma_{el}$, where the collisional interaction rate $d\Gamma_{el}$ is~\cite{GLV,Djordjevic2006} (note $d^{3} N_{J} = d_{R} |J(p^\prime)|^{2} \frac{d^{3}\vec{\mathbf{p^\prime}}}{( 2\pi )^{3} 2 E^\prime} $, 
with $d_R=3$)
\beqar 
d^{3}N_J \; d\Gamma_{el} \; &\approx& \frac{1}{2} \sum_{spins} |M_{el}|^{2} \frac{d^{3} \vec{{\bf p^\prime}}}{(2 \pi)^{3} 2 E^\prime } \frac{d^{3} \vec{{\bf k}}}{(2 \pi)^{3} 2 |\vec{\mathbf{k}}|} \frac{d^{3} \vec{{\bf k^\prime}}}{(2 \pi)^{3} 2 |\vec{\mathbf{k^\prime}}|} \nonumber \\
&&\times\sum_{\xi=q, \bar{q}, g} n_{eq}^\xi (|\vec{\mathbf{k}}|) [1\pm n_{eq}^\xi (|\vec{\mathbf{k^\prime}}|)].
\eeqar{imm1a} 
Following Refs.~\cite{Djordjevic2006,BT_fermions}, the $\pm n^{\xi}_{\rm eq}(|\vec{\mathbf{k'}}|)$ term does not contribute to the collisional energy loss, so we keep only $n_{\rm eq}(|\vec{\mathbf{k}}|)$:
\beqar 
d^{3}\!N_J  d\Gamma_{el} \!\approx \! &&\frac{1}{2} \!\sum_{spins}\! |M_{el}|^{2} \!\frac{d^{3} \vec{{\bf p^\prime}}}{(2 \pi)^{3} 2 E^\prime } \frac{d^{3} \vec{{\bf k}}}{(2 \pi)^{3} 2|\vec{\mathbf{ k}}| } \frac{d^{3} \vec{{\bf k^\prime}}}{(2 \pi)^{3} 2 |\vec{\mathbf{k^\prime}}| }  \, \hspace{1.2em} \nonumber \\
&& \times n_{eq} (|\vec{\mathbf{k}}|),
\eeqar{imm1aELoss} 
where $n_{eq}(|\vec{\mathbf{k}}|)=\sum_{\xi=q, \bar{q}, g} n_{eq}^\xi (|\vec{\mathbf{k}}|)  =\frac{N}{e^{|\vec{\mathbf{k}}|/T}-1}+\frac{N_f}{e^{|\vec{\mathbf{k}}|/T}+1}$ is the equilibrium momentum distribution at temperature $T$ including quark, antiquark and gluon contributions ($N$ and $N_f$ denote numbers of colors and flavors, respectively).

The collisional energy loss of a jet with cone radius $R$ is obtained from the
elastic rate by weighting with the energy transfer $\omega$, averaging over the
direction of the jet velocity $\vec{\mathbf{v}}$, and then restricting the
result to the energy that is transported outside a cone of opening angle $R$
around the jet direction:
\beqar
  \Delta E_{\rm el}^{(R)}
  &=& \frac{C_R}{E^2} \int \frac{d^3\vec{\mathbf{k}}}{(2\pi)^3\,2|\vec{\mathbf{k}}|}\, n_{\rm eq}(|\vec{\mathbf{k}}|)
      \int \frac{d^3\vec{\mathbf{k'}}}{(2\pi)^3\,2|\vec{\mathbf{k'}}|}\;
      \omega\, \nonumber \\
      && \times \Big\langle \frac{\sin^2\left((\omega-\vec{\mathbf{v}}\cdot\vec{\mathbf{q}})\frac{L}{2v}\right)}
         {(\omega-\vec{\mathbf{v}}\cdot\vec{\mathbf{q}})^{2}}\,
      \frac{1}{2}\sum_{\rm spins} |{\cal M}|^2 \,  \Big\rangle_{\hat v,\;R}, \hspace{1.4em}\,
\eeqar{eq:DeltaE_R}
where
\begin{widetext}
\beqar
\frac{1}{2}\sum_{\rm spins}\,|{\cal M}|^2
&=& 16\,g^4 E^2 \Bigg\{
|\Delta_L(q)|^2\Big(|\vec{\mathbf{k}}|\,|\vec{\mathbf{k^\prime}}|
+ (\vec{\mathbf{k}}\!\cdot\!\vec{\mathbf{k^\prime}})\Big)
+ 2\,\mathrm{Re} \big(\Delta_L(q)\Delta_T(q)^*\big)
\Big[\,|\vec{\mathbf{k}}|\, v_i P_{ij} k'_j
     + |\vec{\mathbf{k^\prime}}|\, v_i P_{ij} k_j \Big]
\nonumber\\
&& \hspace{-1.2cm}+ |\Delta_T(q)|^2 \Bigg[
2\left(\vec{\mathbf{v}}\!\cdot\!\vec{\mathbf{k}}
-\frac{(\vec{\mathbf{v}}\!\cdot\!\vec{\mathbf{q}})(\vec{\mathbf{q}}\!\cdot\!\vec{\mathbf{k}})}{|\vec{\mathbf{q}}|^2}\right)
\left(\vec{\mathbf{v}}\!\cdot\!\vec{\mathbf{k^\prime}}
-\frac{(\vec{\mathbf{v}}\!\cdot\!\vec{\mathbf{q}})(\vec{\mathbf{q}}\!\cdot\!\vec{\mathbf{k^\prime}})}{|\vec{\mathbf{q}}|^2}\right)
+\Big(|\vec{\mathbf{k}}|\,|\vec{\mathbf{k^\prime}}|
-\vec{\mathbf{k}}\!\cdot\!\vec{\mathbf{k^\prime}}\Big)
\left(v^2-\frac{(\vec{\mathbf{v}}\!\cdot\!\vec{\mathbf{q}})(\vec{\mathbf{q}}\!\cdot\!\vec{\mathbf{v}})}{|\vec{\mathbf{q}}|^2}\right)
\Bigg]\Bigg\}\,. \hspace{0.65cm}
\eeqar{Cal_M_Sq}
\end{widetext}
To compute the jet energy loss $\Delta E_{\rm el}^{(R)}$, we first need to
evaluate the following integrals, where $\alpha$ is the angle between the
recoil direction and the jet axis, and $\Theta(\alpha-R)$ selects recoils
outside the cone of opening angle $R$. Here $R$ denotes the jet-cone radius around the jet axis; for high-$p_T$ jets at midrapidity this can be directly mapped onto the standard jet-radius parameter used in experimental analyses.

\begin{widetext}
\beqar
{\cal J}_1^{(R)} &=& \int \frac {d \Omega}{ 4 \pi } \; 
\frac{\sin^2\left((\omega- \vec{\mathbf{v}}\cdot\vec{\mathbf{q}})\frac{L}{2 v}\right)}
{(\omega- \vec{\mathbf{v}}\cdot \vec{\mathbf{q}})^2} \; \Theta(\alpha-R)
=\frac{L}{4 |\vec{\mathbf{q}}| v^2}
\left[\mathrm{Si}\left( (\omega+v |\vec{\mathbf{q}}|)\frac{L}{v}\right)-
\mathrm{Si}\left((\omega- v |\vec{\mathbf{q}}|\cos R)\frac{L}{v}\right)\right] \nonumber \\ &-& 
\frac{1}{4 v |\vec{\mathbf{q}}|} \left[
\frac{1-\cos\left((\omega+v |\vec{\mathbf{q}}|)\frac{L}{v}\right)}
{\omega+v |\vec{\mathbf{q}}|}-\frac{1-\cos\left((\omega-v |\vec{\mathbf{q}}|\cos R)\frac{L}{v}\right)}
{\omega-v |\vec{\mathbf{q}}|\cos R} \right] \; ,
\eeqar{J1}
\beqar
{\cal J}_2^{(R)} &=& \int \frac {d \Omega}{ 4 \pi } \; 
\frac{\sin^2\left((\omega- \vec{\mathbf{v}}\cdot\vec{\mathbf{q}})\frac{L}{2 v}\right)}
{(\omega- \vec{\mathbf{v}}\cdot \vec{\mathbf{q}})^2} \; (\omega- \vec{\mathbf{v}}\cdot \vec{\mathbf{q}})\; \Theta(\alpha-R)
\nonumber \\
&=& \frac{1}{4 v |\vec{\mathbf{q}}|} 
\left[\mathrm{Ci}\left((\omega-v |\vec{\mathbf{q}}|\,\cos R)\frac{L }{v}\right)-
\mathrm{Ci}\left( (\omega+v |\vec{\mathbf{q}}|)\frac{L}{v}\right)+
\ln\left|\frac{\omega+v |\vec{\mathbf{q}}|}{\omega-v |\vec{\mathbf{q}}|\,\cos R}\right| \right] \; ,
\eeqar{J2}
\beqar
{\cal J}_3^{(R)}  &=& \int \frac {d \Omega}{ 4 \pi } \; 
\frac{\sin^2\left((\omega- \vec{\mathbf{v}}\cdot\vec{\mathbf{q}})\frac{L}{2 v}\right)}
{(\omega- \vec{\mathbf{v}}\cdot \vec{\mathbf{q}})^2} \; (\omega- \vec{\mathbf{v}}\cdot \vec{\mathbf{q}})^2\; \Theta(\alpha-R)
=\frac{1+\cos R}{4}
+\frac{
\sin\!\left((\omega-v |\vec{\mathbf{q}}|\,\cos R)\frac{L }{v}\right)
-\sin\!\left((\omega+v |\vec{\mathbf{q}}|)\frac{L}{v}\right)}{4\,L\,|\vec{\mathbf{q}}|}. \hspace{0.6cm} 
\eeqar{J3}
By inserting Eq.~(\ref{Cal_M_Sq}) into Eq.~(\ref{eq:DeltaE_R}), using
Eqs.~(\ref{J1})--(\ref{J3}), and carrying out the angular algebra in close
analogy to Ref.~\cite{Djordjevic2006}---now extended to the cone-restricted
integrals ${\cal J}^{(R)}_{1,2,3}$---one arrives at
\beqar
\Delta E_{el}^{(R)} 
&=& 
\int_0^\infty n_{eq}(|\vec{\mathbf{k}}|) d |\vec{\mathbf{k}}| \; 
\left( \int_0^{|\vec{\mathbf{k}}|} |\vec{\mathbf{q}}|d |\vec{\mathbf{q}}| 
\int_{-|\vec{\mathbf{q}}|}^{|\vec{\mathbf{q}}|}\; \omega d \omega \;+
\int_{|\vec{\mathbf{k}}|}^{|\vec{\mathbf{q}}|_{max}} |\vec{\mathbf{q}}|d |\vec{\mathbf{q}}| 
\int_{|\vec{\mathbf{q}}|-2|\vec{\mathbf{k}}| }^{|\vec{\mathbf{q}}|}\; 
\omega d \omega \; \right) \frac{8 \,C_R \alpha_s^2}{\pi^2}  
 \nonumber \\
&& \hspace*{-1.5cm} \Biggl\{ |\Delta_L(q)|^2 \frac{(2 |\vec{\mathbf{k}}|+\omega)^2  - 
|\vec{\mathbf{q}}|^2}{2} {\cal J}_1^{(R)} +
|\Delta_T(q)|^2 \frac{(|\vec{\mathbf{q}}|^2-\omega^2)
[(2 |\vec{\mathbf{k}}|+\omega)^2+ |\vec{\mathbf{q}}|^2]}
{4 |\vec{\mathbf{q}}|^4} \left[(v^2 |\vec{\mathbf{q}}|^2-\omega^2) {\cal J}_1^{(R)} +
2 \omega {\cal J}_2^{(R)}- {\cal J}_3^{(R)}\right] \Biggr\} , \hspace{0.6cm}
\eeqar{Eel_result_finiteL_R} 
\end{widetext}
with $C_R=4/3$, $|\vec{\mathbf{q}}|_{max}=\text{Min}[E, \frac{2 |\vec{\mathbf{k}}| 
(1-|\vec{\mathbf{k}}|/E)}{1-v+2|\vec{\mathbf{k}}|/E}]$~\cite{Djordjevic2006,TG}.

\subsection{Infinite--medium limit}
In the case of an infinite QCD medium, the collisional energy loss per unit 
length $\frac{d E_{el}}{d L}$ is computed by assuming that the jet is produced 
at $x_0=-\infty$. For phenomenological comparison one then approximates the finite-size loss as $(dE_{\rm el}/dL)\,L$.

The jet collisional energy loss per unit path length is obtained by multiplying
Eq.~\eqref{Eel_result_finiteL_R} by $2v/(\pi L)$ and taking the limit
$L\to\infty$. In this limit one has the standard representation of the
$\delta$–function,
\begin{equation}
  \frac{2v}{\pi L}\,
  \frac{\sin^2\left((\omega-\mathbf{v}\!\cdot\!\mathbf{q})L/(2v)\right)}
       {(\omega-\mathbf{v}\!\cdot\!\mathbf{q})^{2}}
  \xrightarrow[L\to\infty]{}\;
  \delta(\omega-\mathbf{v}\!\cdot\!\mathbf{q})\,,
  \label{eq:delta_limit}
\end{equation}
and, correspondingly, $J_2^{(R)}$ and $J_3^{(R)}$
vanish, while $J_1^{(R)}$ reduces to a simple step–function kernel enforcing
$-v |\vec{\mathbf{q}}|<\omega < v|\mathbf{q}|\cos R$. Using these limiting forms in 
Eq.~\eqref{Eel_result_finiteL_R} and performing the remaining integrations,
one finds that the jet collisional energy loss per unit length in an infinite QCD 
medium reduces to ($C_R=4/3$)
\begin{widetext}
\beqar
\frac{d E_{el}^{(R)}}{d L} &=&
\int_0^\infty n_{eq}(|\vec{\mathbf{k}}|) d |\vec{\mathbf{k}}| \; 
\left( \int_0^{|\vec{\mathbf{k}}|/(1+v)} d |\vec{\mathbf{q}}| 
\int_{-v |\vec{\mathbf{q}}|}^{v |\vec{\mathbf{q}}|\cos R}\; \omega d \omega \;+ 
\int_{|\vec{\mathbf{k}}|/(1+v)}^{|\vec{\mathbf{q}}|_{max}} d |\vec{\mathbf{q}}|
\int_{|\vec{\mathbf{q}}|-2|\vec{\mathbf{k}}| }^{\omega_{max}^{R}}\; 
\omega d \omega \; \right) \frac{2 C_R \alpha_s^2}{\pi \, v^2 }  
\nonumber \\
&& \hspace*{0.5cm}\Biggl\{ |\Delta_L(q)|^2 \frac{(2 |\vec{\mathbf{k}}|+\omega)^2  - 
|\vec{\mathbf{q}}|^2}{2}  + |\Delta_T(q)|^2 \frac{(|\vec{\mathbf{q}}|^2-\omega^2)
[(2 |\vec{\mathbf{k}}|+\omega)^2+ |\vec{\mathbf{q}}|^2]}
{4 |\vec{\mathbf{q}}|^4} (v^2 |\vec{\mathbf{q}}|^2-\omega^2) \Biggr\}, 
\eeqar{Eel_limit_Final}
where $\omega_{max}^{R}=\text{Max}[|\vec{\mathbf{q}}|-2|\vec{\mathbf{k}}|,v |\vec{\mathbf{q}}|\cos R]$. We note that, in the limit $R\to 0$, both Eq.~\ref{Eel_result_finiteL_R} and 
Eq.~\ref{Eel_limit_Final} reduce to the parton collisional energy-loss expressions given by
Eqs.~(14) and~(16) in~\cite{Djordjevic2006,TG}, providing a consistency check
of our previous finite-size calculation of parton collisional energy loss.
\end{widetext}
\subsection{Beyond the strict HTL limit}
The analytical derivation above is performed within the pure HTL framework and assumes a fixed coupling.
In the numerical implementation, however, we incorporate two standard improvements.
First, the electric screening (Debye) mass $\mu_E(T)$ is determined self-consistently by solving the equation of Ref.~\cite{Peshier:2006ah},
\beqar
\frac{\mu^2_E(T)}{\Lambda_{QCD}^2}\ln\!\left(\frac{\mu^2_E(T)}{\Lambda_{QCD}^2}\right)
=\frac{1+N_f/6}{11-(2/3)\,N_f}\left(\frac{4\pi T}{\Lambda_{QCD}}\right)^2,
\eeqar{mu}
which can be written in closed form using Lambert's $W$ function,
\beqar
\mu_E\!=\!\sqrt{\Lambda_{QCD}^2\frac{\psi(T)}{W(\psi(T))}}, 
\psi(T)\!=\!\frac{1\!+\!N_f/6}{11\!-\!(2/3)N_f}\!\left(\!\frac{4\pi T}{\Lambda_{QCD}}\!\right)^2\!.\! \hspace{0.6cm}
\eeqar{Debye_mass}
Second, instead of a constant coupling we use the one-loop running coupling
\beqar
\alpha_S(Q^2)=\frac{4\pi}{(11-(2/3)\,N_f)\,\ln(Q^2/\Lambda_{QCD}^2)},
\eeqar{alpha}
where the coupling factors entering the collisional energy-loss expression are evaluated as
$\alpha_S(E\,T)\,\alpha_S(\mu_E^2)$~\cite{PP_PRD,MD_PLB}.

\section{Numerical results}
\label{sec:numerical}

In this section we present numerical results for the cone-restricted collisional energy loss $\Delta E^{(R)}_{\rm el}$, i.e., the elastic energy transported outside a jet cone of radius $R$.
We consider a QGP with $\Lambda_{\rm QCD}=0.2$, $N_f=3$, and $T=0.335~{\rm GeV}$ for 0--10\% central Pb+Pb collisions at $\sqrt{s_{\rm NN}}=5.02~{\rm TeV}$~\cite{Zigic:2018smz}, and use $M_c=1.2~{\rm GeV}$ and $M_b=4.75~{\rm GeV}$ for heavy quarks, while $M_l=\mu_E/\sqrt{6}$ for light quarks.

\paragraph{Cone dependence and comparison to radiative energy loss.}

\begin{figure}[htbp]
  \centering
  \includegraphics[scale=0.6]{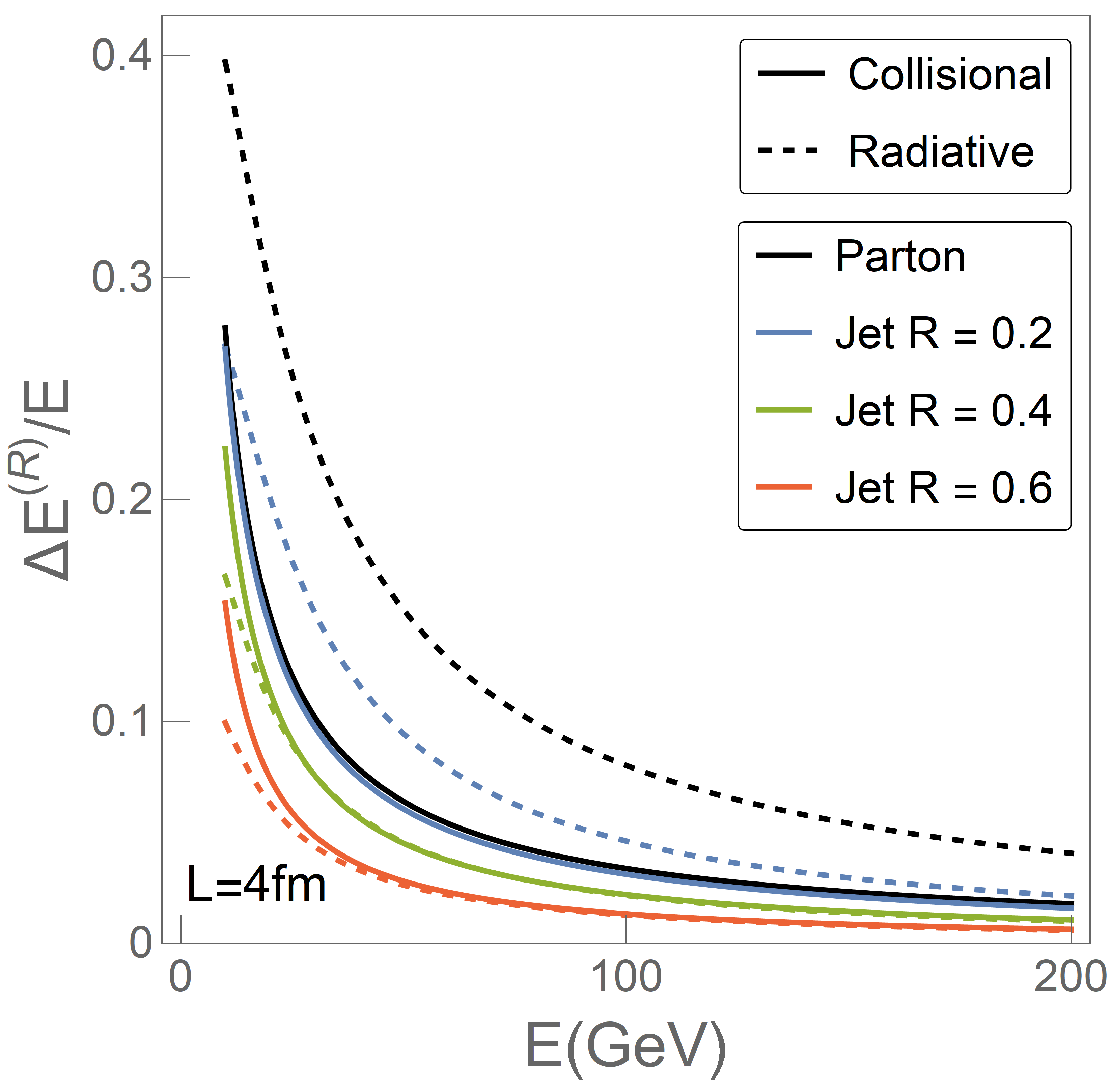}
  \vspace{-0.7em}
  \caption{Fractional energy loss $\Delta E^{(R)}/E$ of light-quark jets as a function of jet (parton) energy $E$ for a medium path length $L=4~\mathrm{fm}$. Solid curves show the collisional contribution and dashed curves show the radiative contribution~\cite{Karmakar:2024fkn}. Results are shown for a parton ($R=0$) and for  jets with cone radii $R=0.2$, $0.4$, and $0.6$, as indicated in the legend.}
\label{fig:C1}
\end{figure}

Figure~\ref{fig:C1} shows the fractional energy loss $\Delta E^{(R)}/E$ of light-quark jets as a function of jet
(parton) energy for a fixed path length $L=4~{\rm fm}$.
For comparison, radiative out-of-cone losses are evaluated using the formalism of Ref.~\cite{Karmakar:2024fkn} (dashed curves) with the same medium parameters and kinematic setup as used in the present work, while our collisional
results are shown by solid curves.
As expected, in the partonic case ($R=0$) the radiative component is substantially larger than the
collisional one throughout the plotted energy range.
However, the cone dependence is qualitatively different for the two mechanisms:
the radiative out-of-cone loss decreases in a systematic manner with increasing $R$, whereas the
collisional loss exhibits a delayed onset, with $\Delta E^{(R)}_{\rm el}$ for a narrow jet ($R=0.2$)
remaining very close to the partonic ($R=0$) result, followed by a noticeably stronger reduction for
$R=0.4$ and an even larger reduction for $R=0.6$.
This difference in the $R$-dependence has an important consequence: while radiative loss dominates
for $R=0$, the relative importance of the collisional component increases with $R$, and for the
widest cone shown ($R=0.6$) the collisional out-of-cone energy loss becomes the dominant contribution in
Fig.~\ref{fig:C1}.
Therefore, for  jets the common partonic hierarchy ``radiative $\gg$ collisional'' need
not hold, and the elastic component can become comparable to, or larger than, the radiative one
depending on the jet radius.

\begin{figure*}[htbp]
  \centering
  \includegraphics[scale=0.57]{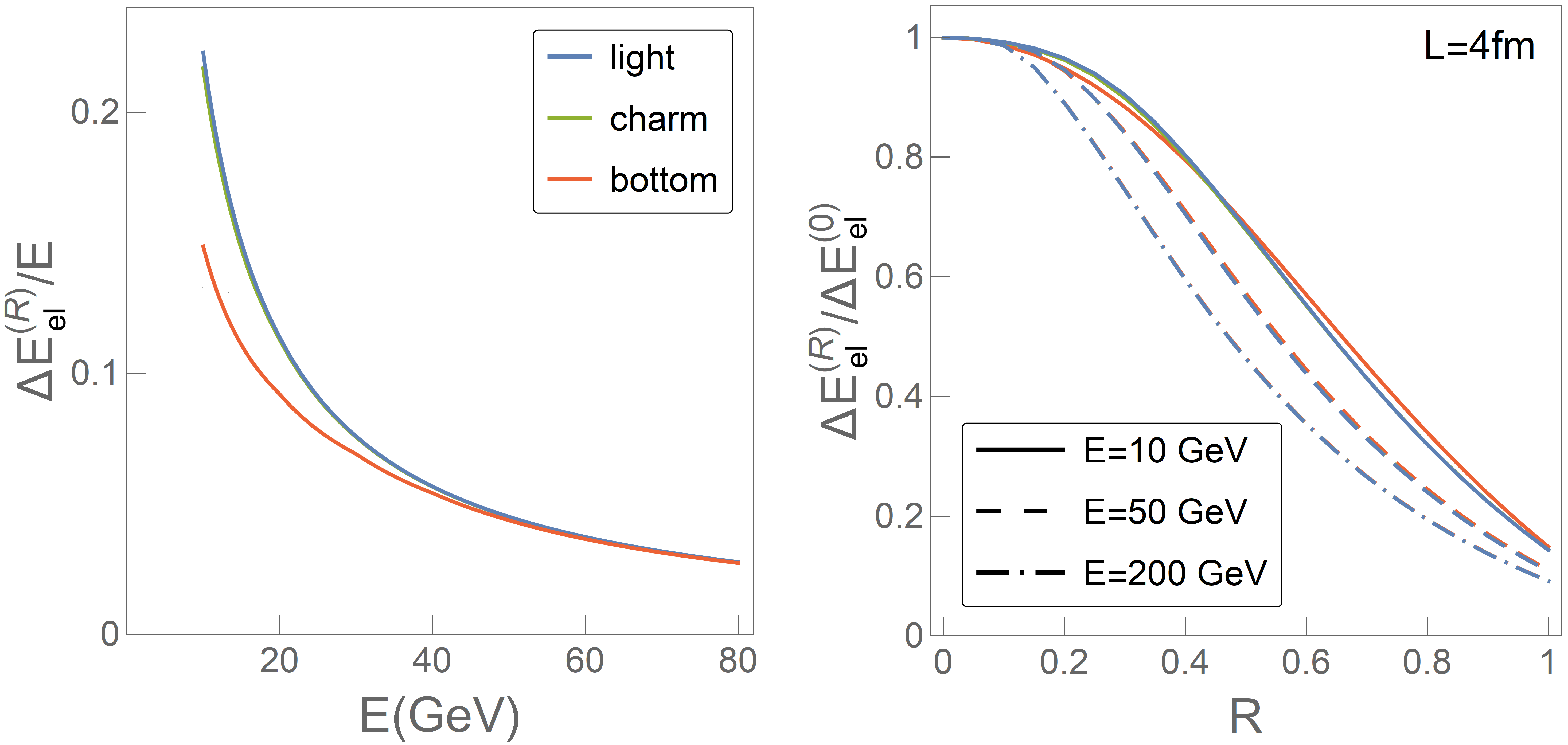}
  \vspace{-0.7em}
\caption{Collisional fractional energy loss and its cone dependence for light-, charm-, and bottom-quark--initiated jets at $L=4~\mathrm{fm}$. Left panel: $\Delta E^{(R)}_{el}/E$ as a function of $E$ at fixed cone radius $R=0.4$. Right panel: ratio $\Delta E^{(R)}_{el}/\Delta E^{(0)}_{el}$ as a function of the jet cone radius $R$ at fixed energies $E=10$, $50$, and $200~\mathrm{GeV}$ (line styles), shown for light-, charm-, and bottom-quark--initiated jets (colors).}
\label{fig:C2}
\end{figure*}

\paragraph{Light vs.\ heavy flavor jets.}
Figure~\ref{fig:C2} (left) compares $\Delta E^{(R)}_{el}/E$ at fixed $R=0.4$ for light, charm and bottom jets.
A clear mass hierarchy is present at low energies: bottom jets lose less energy than charm and
light jets up to $E\sim \mathcal{O}(40~{\rm GeV})$, while at higher energies the three curves approach each
other as expected when mass effects become subleading.
Figure~\ref{fig:C2} (right) isolates the cone effect by plotting the ratio $\Delta E^{(R)}_{el}/\Delta E^{(0)}_{el}$ as a
function of $R$ for $E=10$, $50$ and $200~{\rm GeV}$.
For each energy, the ratios for light, charm and bottom nearly coincide, indicating that the
cone dependence of the collisional component is only weakly sensitive to the quark mass.

The $R$ dependence is highly non-linear: the ratio $\Delta E^{(R)}_{\rm el}/\Delta E^{(0)}_{\rm el}$ stays close to unity for small $R$, drops rapidly at intermediate $R$, and saturates at large $R$, with the onset shifting to smaller $R$ as $E$ increases.
The weak variation at small $R$ is expected since the cone dependence enters the finite-size kernels through $\cos R$ (e.g.\ via $\omega - v|{\bf q}|\cos R$), so that
$\Delta E^{(R)}_{\rm el}=\Delta E^{(0)}_{\rm el}+\mathcal{O}(1-\cos R)\simeq \Delta E^{(0)}_{\rm el}+\mathcal{O}(R^2)$.
Even at $R=1$ the ratio remains at the $\mathcal{O}(10\%)$ level, indicating a non-negligible large-angle component of elastic energy transport.
Thus neither limiting treatment often assumed for the elastic component---$\Delta E^{(R)}_{\rm el}\!\approx\!\Delta E^{(0)}_{\rm el}$ or $\Delta E^{(R)}_{\rm el}\!\approx\!0$---is generically justified.

\paragraph{Path-length dependence and finite-size effects.}
\begin{figure}[htbp]
  \centering
  \includegraphics[scale=0.23]{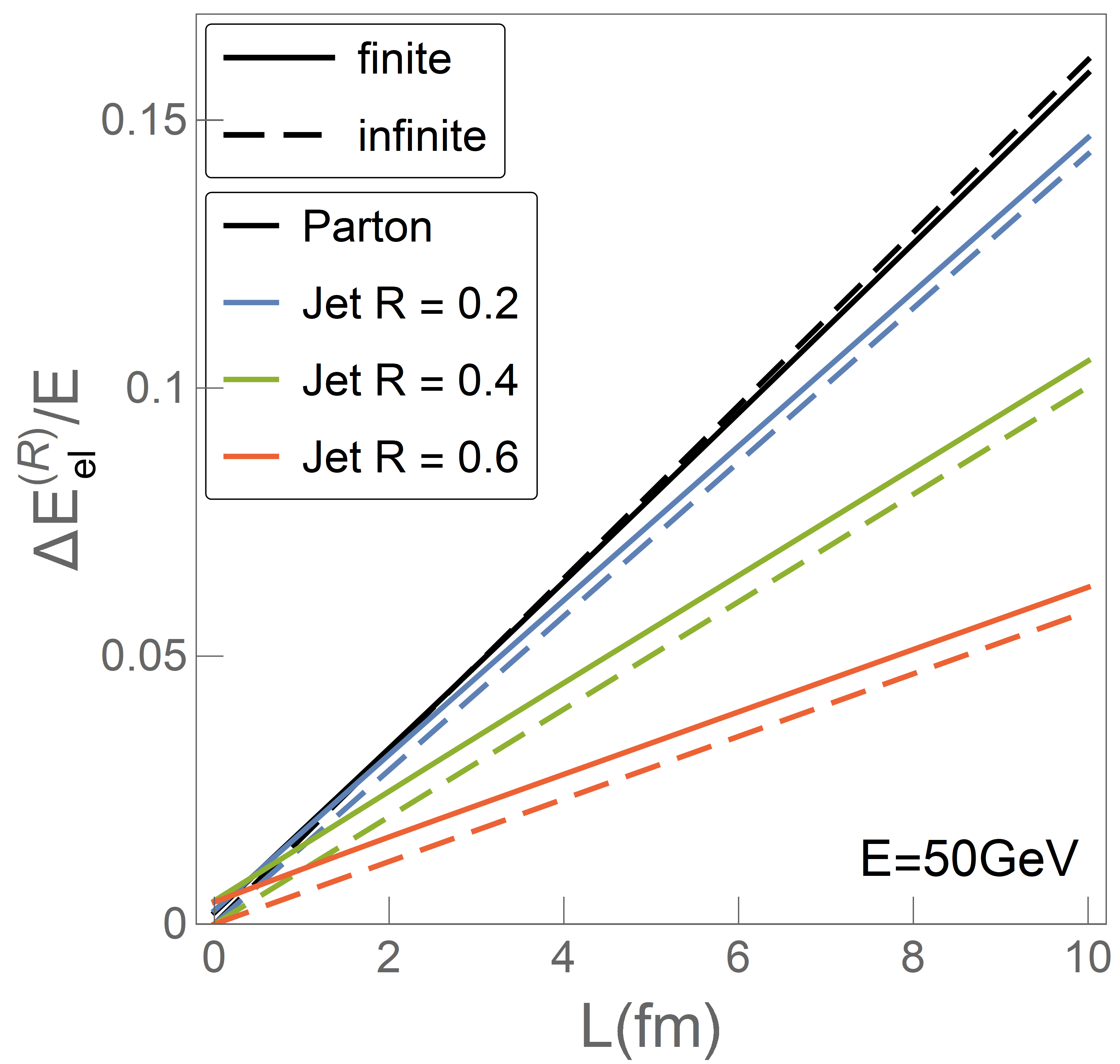}
  \vspace{-0.7em}
\caption{Collisional fractional energy loss $\Delta E^{(R)}_{el}/E$ as a function of path length $L$ for light-quark--initiated jets with $E=50~\mathrm{GeV}$. Solid curves correspond to the finite-size calculation and dashed curves to the infinite-medium (large-$L$) limit. Results are shown for a parton ($R=0$) and for jets with cone radii $R=0.2$, $0.4$, and $0.6$, as indicated in the legend.}
\label{fig:C3}
\end{figure}

Figure~\ref{fig:C3} shows $\Delta E^{(R)}_{el}/E$ for a $50~{\rm GeV}$ jet as a function of the path length $L$, and
compares the finite-size calculation (solid) with the infinite-medium approximation (dashed).
For all cone radii, the energy loss remains very close to linear in $L$, with finite-size effects producing only modest quantitative corrections relative to the infinite-medium approximation.
This indicates that the main impact of the jet cone is to reduce the overall magnitude of the out-of-cone energy transfer,
without inducing a qualitatively new path-length scaling.

This behavior is in marked contrast to our radiative out-of-cone results, where increasing the jet radius reduces the effective path-length dependence of the radiative jet energy loss (see Ref.~\cite{Karmakar:2024fkn} and
references therein, as well as related discussions on cone-size dependence of jet suppression~\cite{MehtarTani:2021dqn,Wu:2023PRC}).
The approximately linear-in-$L$ scaling of the collisional out-of-cone component for all $R$, together with the weakening of the radiative $L$ dependence at larger $R$, naturally explains Fig.~\ref{fig:C1}: as the cone opens, the relative importance of elastic energy transfer increases and can become dominant for sufficiently large jet radii.

\paragraph{Medium response contribution.}
\begin{figure*}[htbp]
  \centering
  \includegraphics[scale=0.56]{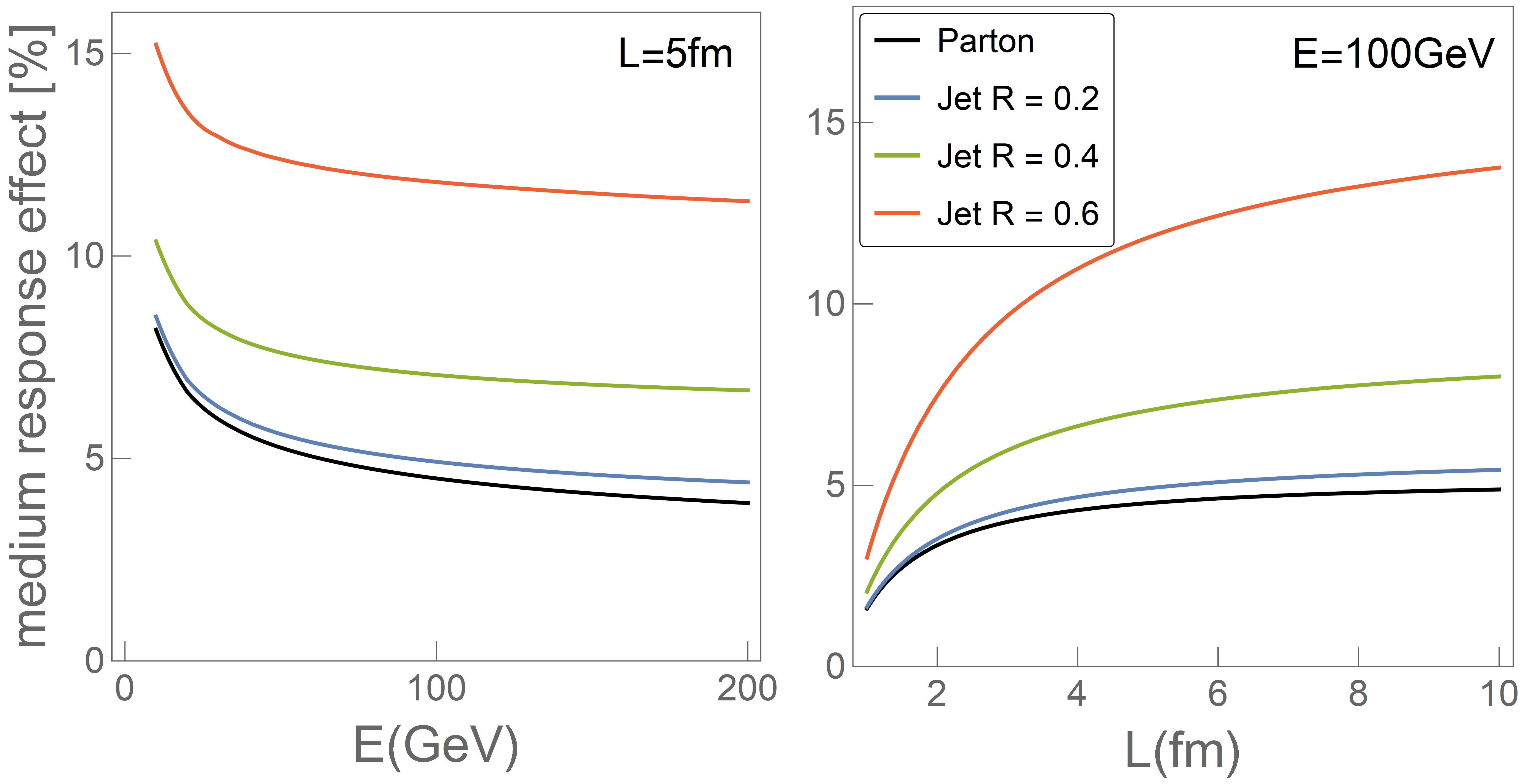}
  \vspace{-0.7em}
\caption{Medium-response effect (in percent) for collisional energy loss of light-quark--initiated jets, defined as
$\frac{\Delta E^{(R)}_{\rm el,no\,resp}-\Delta E^{(R)}_{\rm el}}{\Delta E^{(R)}_{\rm el}}$,
where $\Delta E^{(R)}_{\rm el}$ denotes the collisional energy loss including medium response.
Left: medium-response effect as a function of jet (parton) energy $E$ at fixed path length $L=5~\mathrm{fm}$.
Right: medium-response effect as a function of path length $L$ at fixed energy $E=100~\mathrm{GeV}$.
Curves correspond to the light-quark parton limit ($R=0$) and to jets with cone radii $R=0.2$, $0.4$, and $0.6$, as indicated in the legend.}
\label{fig:C4}
\end{figure*}
Finally, Fig.~\ref{fig:C4} quantifies the medium-response effect directly within our analytic, HTL-resummed finite-size expression. In our leading-order elastic framework, the ``medium response'' refers to the $\omega<0$ (energy-gain) sector associated with thermal recoil/detailed-balance contributions.
Specifically, since the energy-transfer integral in our formulation retains the $\omega<0$ (energy-gain) contribution associated with medium response,
we can evaluate both the full result $\Delta E^{(R)}_{\rm el}$ and, within the same microscopic framework, a ``no response'' variant
$\Delta E^{(R)}_{\rm el,no\,resp}$ obtained by excluding the $\omega<0$ sector (from Eq.~(\ref{Eel_result_finiteL_R})), thereby providing a controlled and unambiguous assessment of the importance of medium response for jets.

We find that for partons ($R=0$) the effect is small, at the level of a few percent, but it increases systematically
with jet radius and reaches $\sim 15\%$ for $R=0.6$ in the plotted kinematics.
The medium-response correction decreases with increasing energy (left panel) and increases with path length (right panel), and in both cases shows
signs of saturation.
These results demonstrate that medium response can provide a non-negligible, $R$-dependent, correction to jet collisional energy loss; for realistic jet radii it can exceed the finite-size correction to the elastic out-of-cone loss.

\section{Conclusion}

We derived a compact HTL-resummed expression for the leading-order jet collisional energy loss in a finite-size QCD matter with explicit cone selection.
Defining the jet energy inside a cone of radius $R$ yields the corresponding out-of-cone elastic loss with an explicit separation between the contribution of the primary (jet-initiating) parton and that associated with recoiling medium partons, interpolating between the partonic limit ($R\!\to\!0$) and complete recovery ($R\!\to\!\pi$) for both light and heavy flavor.

Numerically, the elastic out-of-cone component exhibits a pronounced non-linear $R$ dependence that differs from the radiative baseline, so its relative importance increases with $R$ and can become comparable to or exceed the radiative contribution for sufficiently large cones.
The cone dependence is only weakly sensitive to the quark mass, while the path-length scaling remains close to linear for all $R$ with only modest finite-size corrections; together with the reduced radiative $L$ dependence at larger $R$, this explains the increasing relative role of elastic energy transfer with jet radius.
Finally, isolating the $\omega<0$ sector provides a controlled estimate of medium response, which is a few-percent effect for partons but can reach the $\mathcal{O}(10\%)$ level for realistic jet radii.

These results provide a compact analytic baseline for jet collisional energy loss in finite-size QCD matter and demonstrate that neither extreme assumption---fully out-of-cone loss nor full in-cone recovery of the elastic component---is generally justified.

\section*{Acknowledgments} This work is supported by the Ministry of Science and Technological Development of the Republic of Serbia and by the Serbian Academy of Sciences and Arts.

\end{document}